\documentclass[runningheads,11pt]{article}

\usepackage[margin=1in, letterpaper]{geometry}
\usepackage{multirow}

\usepackage{algorithm2e}

\usepackage{amsthm}

\newtheorem{question}{Open Question}

\usepackage{wrapfig}

\usepackage{amsmath,amssymb,graphicx}
\usepackage{verbatim}
\usepackage{centernot,cancel}
\usepackage{xcolor}
\usepackage[colorinlistoftodos,prependcaption,textsize=tiny]{todonotes}

\usepackage{authblk}
\usepackage{xspace}
\usepackage{xcolor}
\usepackage{makecell}
\usepackage{url}
\usepackage{hyperref}
\usepackage{floatrow}

\usepackage{subcaption}
\usepackage{graphicx}
\graphicspath{ {./images/} }

\newcommand{\argmin}{\ensuremath{\text{argmin}}\xspace}

\newcommand{\erfc}{\ensuremath{\text{erfc}}\xspace}

\newcommand{\minenergy}{\textsc{MinEnergy}\xspace}
\newcommand{\minstructure}{\textsc{MinStructure}\xspace}
\newcommand{\avgenergy}{\textsc{AvgEnergy}\xspace}
\newcommand{\avgstructure}{\textsc{AvgStructure}\xspace}

\newcommand{\srtio}{\ensuremath{\text{SrTiO}_3}\xspace}

\newcommand{\ytio}{\ensuremath{\text{Y}_2\text{Ti}_2\text{O}_7}\xspace}

\newcommand{\mgtio}{\ensuremath{\text{MgTiO}_3}\xspace}

\newcommand{\dist}{\ensuremath{\texttt{d}}\xspace}

\newcommand{\angstrom}{\mbox{\normalfont\AA}}

\newcommand{\R}{\mathbb{R}}

\usepackage{tcolorbox}
\usepackage{amsfonts}

\newcount\Comments  % 0 suppresses notes to selves in text
\definecolor{darkgreen}{rgb}{0,0.6,0}
\newcommand{\kibitz}[2]{\ifnum\Comments=1{\color{#1}{#2}}\fi}

\title{Crystal Structure Prediction via Oblivious Local Search
}

\author{
Dmytro Antypov$^{1,2}$, Argyrios Deligkas$^{3}$, Vladimir Gusev$^{2,4}$, Matthew J. Rosseinsky$^{1}$, Paul G. Spirakis$^{4,5}$, and Michail Theofilatos$^{2,4}$
}

\date{%
    $^1$Department of Chemistry, University of Liverpool, UK\\
    $^2$Leverhulme Research Centre for Functional Materials Design, UK \\
    $^3$Department of Computer Science, Royal Holloway University of London, UK\\%
    $^4$Department of Computer Science, University of Liverpool, UK\\%
    $^5$Computer Engineering and Informatics Department, University of Patras, Greece\\[1ex]%
    Email: \{D.Antypov,  Vladimir.Gusev, M.J.Rosseinsky, P.Spirakis, Michail.Theofilatos\}@liverpool.ac.uk, Argyrios.Deligkas@rhul.ac.uk
}

\begin{document}

\maketitle

%\titlerunning{}

%\usepackage{fancyhdr}
%\fancyhf{}
%\pagestyle{fancy}
%\lhead{\centering{\textbf{APPENDIX}}}
%\cfoot{\thepage}

\begin{abstract} \normalsize
%In the Crystal Structure Prediction problem the goal is to find a configuration for
%a set of ions in the 3d Euclidean space such that the potential energy of the
%set is minimised. 
%
We study Crystal Structure Prediction, one of the major problems in computational 
chemistry. This is essentially a continuous optimization problem, 
where many different, simple and sophisticated, methods have been proposed and applied. 
%However, many methods are finely tuned to the input composition and involve 
%a large number of parameters to be manually set in advance. 
%Thus, they require expertise in order to be effective. 
%Our goal is to create oblivious algorithms for the Crystal Prediction Problem
%that perform well and require little, or no, knowledge of chemistry.
The simple searching techniques are easy to understand, usually easy to implement, 
but they can be slow in practice. On the other hand, the 
more sophisticated approaches perform well in general, however almost all of them 
have a large number of parameters that require fine tuning and, in the majority of the
cases, chemical expertise is needed in order to properly set them up. In addition, due 
to the chemical expertise involved in the parameter-tuning, these approaches can be 
{\em biased} towards previously-known crystal structures. 
Our contribution is twofold. Firstly, we formalize the Crystal Structure Prediction 
problem, alongside several other intermediate problems, from a theoretical computer
science perspective. Secondly, we propose an oblivious algorithm for Crystal Structure
Prediction that is based on local search. Oblivious means that our algorithm requires 
minimal knowledge about the composition we are trying to compute a crystal structure for. 
In addition, our algorithm can be used as an intermediate step by {\em any} method. 
Our experiments show that our algorithms outperform the standard basin hopping, a well studied algorithm for the problem.
\end{abstract}


\newpage
\section{Introduction}
\label{sec:introduction}
The discovery of new materials has historically been made by experimental investigation guided by chemical understanding. 
This approach can be both time consuming and challenging because of the large space to be explored. %they require excessive amount of labour and fine-tuning expertise.
For example, a ``traditional'' method for discovering 
inorganic solid structures relies on knowledge of crystal chemistry coupled with 
repeating synthesis experiments and systematically varying elemental ratios, each
of which can take lots of time~\cite{combining_magnets,magnetic_oxides}. As a result 
there is a very large unexplored space of chemical systems: only $72\%$ of 
binary systems, $16\%$ of ternary, and just $0.6\%$ of quaternary systems have been
studied experimentally~\cite{villars2001}. 

These inefficiencies forced physical scientists to develop computational approaches in order 
to tackle the problem of finding new materials. The first approach is based on 
data mining where {\em only} pre-existing knowledge is used~\cite{curtarolo2013,gautier2015,hautier2010,nosengo2016,schon2014}. 
Although this approach has proven to be successful, there is the underlying risk of 
missing best-in-class materials by being biased towards {\em known} crystal structures. 
Hence, the second approach tries to fill this gap and aims at finding new materials 
with {\em little, or no,} pre-existing knowledge, by {\em predicting the crystal 
structure} of the material. This approach has led to the discovery of several new, 
counterintuitive, materials whose existence could not be deduced by the structures of  
previously-known  materials~\cite{collins2017}. 

Several heuristic methods have been suggested for crystal structure prediction. 
All these methods are based on the same fundamental principle. 
Every arrangement of ions in the 3-dimensional Euclidean space corresponds to an 
energy value and it defines a point on the {\em potential energy surface}.
%Given a number of 
%ions of some elements, whose charges sum up to zero,
%they define the {\em potential energy surface}; every point on the potential 
%energy surface corresponds to the energy of a 3-dimensional configuration of the ions. 
Then, the crystal structure prediction problem is formulated as a {\em mathematical optimization} 
problem where the goal is to compute the structure that corresponds to the global minimum 
of the potential energy surface, since this is the most likely structure that corresponds to a stable
material. The difficulties in solving this optimization problem is that the potential energy
surface is {\em highly non convex}, with {\em exponentially many}, with respect to the 
number of ions, local minima~\cite{2019review}. For this reason, several different algorithmic techniques were 
proposed ranging from simple techniques, like {\em quasi-random sampling}
\cite{freeman1993,pickard2011,pickard2006,schmidt1996}, {\em basin hopping} 
\cite{goedecker2004,wales1997}, and {\em simulated annealing }
techniques~\cite{pannetier1990,schon1996}, to more sophisticated techniques, like  
{\em evolutionary and genetic algorithms}
\cite{call2007,deaven1995,lonie2011,oganov2006,wang2010}, and {\em tiling} approaches
\cite{collins2017,collins2018}. A recent comprehensive review on these techniques can be found 
in~\cite{2019review}.

The simple searching techniques are easy to understand, usually easy to implement, 
and they are {\em unbiased}, but they can be slow in practice. On the other hand, the 
more sophisticated approaches perform well in general, however almost all of them 
have a large number of parameters that require fine tuning and, in the majority of the
cases, chemical expertise is needed in order to properly set them up. In addition, due 
to the chemical expertise involved in the parameter-tuning, these approaches can be 
{\em biased} towards previously-known structures. 

The majority of the abovementioned heuristic techniques work, at a very high level, in a 
similar way. Given a current solution $x$ for the crystal structure prediction problem, 
i.e., a location for every ion in the 3-dimensional space, they iteratively perform the 
following three steps. 

\begin{enumerate}
    \item \label{step1} Choose a new potential solution $x'$. This can be done by taking into 
    account, or modifying, $x$.
    \item \label{step2} Perform gradient descent on the potential energy surface starting from 
    $x'$, until a local minimum is found. This process is called {\em relaxation} of $x'$.
%    and denoted $\rel(x')$.
    \item \label{step3} Decide whether to keep $x$ as the candidate solution or to update it 
    to the solution found after relaxing $x'$.
\end{enumerate}

For example, basin hopping algorithms randomly choose $x'$, they relax $x'$ and if the energy
of the relaxed structure is lower than the $x$, or a Metropolis criterion is satisfied, 
they accept this as a current solution; else
they keep $x$ and they randomly choose $x''$. The procedure usually stops when the algorithm 
fails to find a structure with lower energy within a predefined number of iterations. The more sophisticated 
algorithms take into account knowledge harvested from chemists and put constraints on the
way $x'$ is selected. For example, the MC-EMMA~\cite{collins2017} and the FUSE~\cite{collins2018}
algorithms use a set of building blocks to construct $x'$. These building blocks are local 
configurations of ions that are present in, or similar to,  known crystal structures. 
%Hence, these algorithms are usually faster in computing known crystal structures, but they
%are implicitly biased towards them, so they can miss other, counterintuitive solutions.
These approaches restrict the search space ,which accelerate search, but reduce the 
number of possible solutions.
%Having a component from a known structure does not mean that adding these will also give you a known structure. It is simpler to say that this restricts the search space which both accelerates search and reduces the number of possible solutions

This general algorithm is easy to understand, however there are some hidden difficulties that make the 
problem more challenging. Firstly, it is not trivial even how to {\em evaluate} the potential energy
of a structure. There are several different methods for calculating the energy of a structure, 
ranging from {\em quantum mechanical} methods, like {\em density functional theory}~\footnote{\url{https://en.wikipedia.org/wiki/Density_functional_theory}}, 
to {\em force fields} methods~\footnote{\url{https://en.wikipedia.org/wiki/Force_field_(chemistry)}}, 
like the {\em Buckingham-Coulomb} potential function. All of which 
though, are hard to compute (see Section~\ref{sec:energy}) from the point of view of (theoretical) computer science and thus only
numerical methods are known and used in practice for them~\cite{gale2003gulp}; still there are cases
where some methods need considerable time to calculate the energy of a structure.
%~\footnote{We have examples of 48 atoms where Barkla supercomputer of Liverpool (105 nodes, 40 cores each) needed two minutes to compute the energy via density function theory methods}. 
This yields another, 
more important, difficulty, the relaxation of a structure. Since it is hard to compute the  
energy of a structure, it is even harder to apply gradient descent on the potential energy surface. For these 
reasons, the majority of the heuristic algorithms depend on {\em external}, well established, codes~\cite{gale2003gulp} 
for computing the aforementioned quantities.
%the energy, with density functional theory or force fields, or perform gradient descent. 
Put differently, both energy computations and relaxations of structures are treated as {\em oracles}
or {\em black boxes}.

\subsection{Our contribution.}
Our contribution is twofold. Firstly, we formalize the Crystal Structure Prediction problem 
from the theoretical computer science perspective; to the best of our knowledge, this is 
among the few papers that attempt to connect computational chemistry and computer science.
En route to this, we introduce several intermediate open problems from computational chemistry 
in CS terms. Any (partial) positive solution to these questions can significantly help 
computational chemists to identify new materials. On the other hand, any negative result can 
formally explain why the discovery of new materials is a notoriously difficult task.

Our second contribution is the partial answer for some of the questions we cast. 
In general, our goal is to create {\em oblivious} algorithms that are easy to implement, 
they are fast, and they work well in practice. With oblivious we mean that we are seeking 
for {\em general procedures}  that require {\em minimal input} and they have zero, or just 
a few, parameters chosen by the user. 
\begin{itemize}
    \item We propose a purely combinatorial method for estimating the energy of a  structure, 
    which we term {\em depth energy computation}. 
    We choose to compare our method against 
GULP~\cite{gale2003gulp}, which is considered to be the state of the art for computing the 
energy of a structure and for performing relaxations when the Buckingham-Coulomb energy is used. 
Our method requires only the charges of the atoms and their corresponding Buckingham coefficients 
to work; see Eq.~\ref{eq:buckingham} in Section~\ref{sec:energy}. 
In addition, it needs only one parameter, the depth $k$. 
We experimentally demonstrate that our method monotonically approximates with respect to $k$ 
the energy computed by GULP and that it achieves an error of $0.0032$ for $k = 6$. 
%Although our naive implementation cannot compete with GULP with respect to CPU time, 
Our experiments show 
that the structure that achieves the minimum energy in depth 1 is likely to be 
the structure with the minimum energy overall. 
In fact we show something much stronger. If the energy of $x$ is lower than the energy of $x'$ 
when it is computed via the depth energy computation for $k = 1$, then, almost always, 
the energy of $x$ will be lower than the energy of $x'$ when it is computed via GULP.
 \item We derive oblivious algorithms for choosing which structure to relax next. All
 of our algorithms are based on local search. More formally, starting with $x$ and using only local 
 changes we select $x'$. We define several ``combinatorial neighborhoods'' and we evaluate their 
 efficiency. Our neighborhoods are oblivious since they only need access to an oracle that 
 calculates the energy of a structure. We show that our method outperforms basin hopping. 
 Moreover, we view our algorithms as an intermediate step before 
 relaxation that can be applied to {\em any} existing algorithm.
 %Our experiments show that the average CPU time needed for a relaxation starting from a point before and after the local search is significantly \argy{is this true, or am I bullshitting here?} decreasing. This is important in practice, since in all of the known algorithms for crystal structure prediction, the relaxation procedure consumes $95\%$ of the CPU time.
\end{itemize}

\section{Preliminaries} \label{sec:preliminaries}

A {\em crystal} is a solid material whose atoms are arranged in a highly ordered 
configuration, forming a {\em crystal structure} that extends in all directions.
A crystal structure is characterized by its {\em unit cell}; a parallelepiped 
that contains atoms in a specific arrangement. 
The unit cell is the {\em period} of the crystal; unit cells are stacked in the 
three dimensional space to form the crystal. 
In this paper we focus on {\em ionically bonded crystals}, which we describe next;
what follows is relevant only on crystals of this type.
In order to fully define the unit cell of a ionically bonded crystal structure, we 
have to specify a {\em composition}, {\em unit cell parameters}, and an {\em arrangement} 
of the ions.

\iffalse
\begin{figure}[h!]
  \centering
    \includegraphics[width=\linewidth]{images/SrTiO3_supercell.pdf}
    \caption{Supercell of \srtio}
    \label{fig:srtio3}
\end{figure}
\fi

\begin{figure}
    \centering
    \begin{subfigure}{.25\textwidth}
        \centering
        \includegraphics[width=1.0\linewidth]{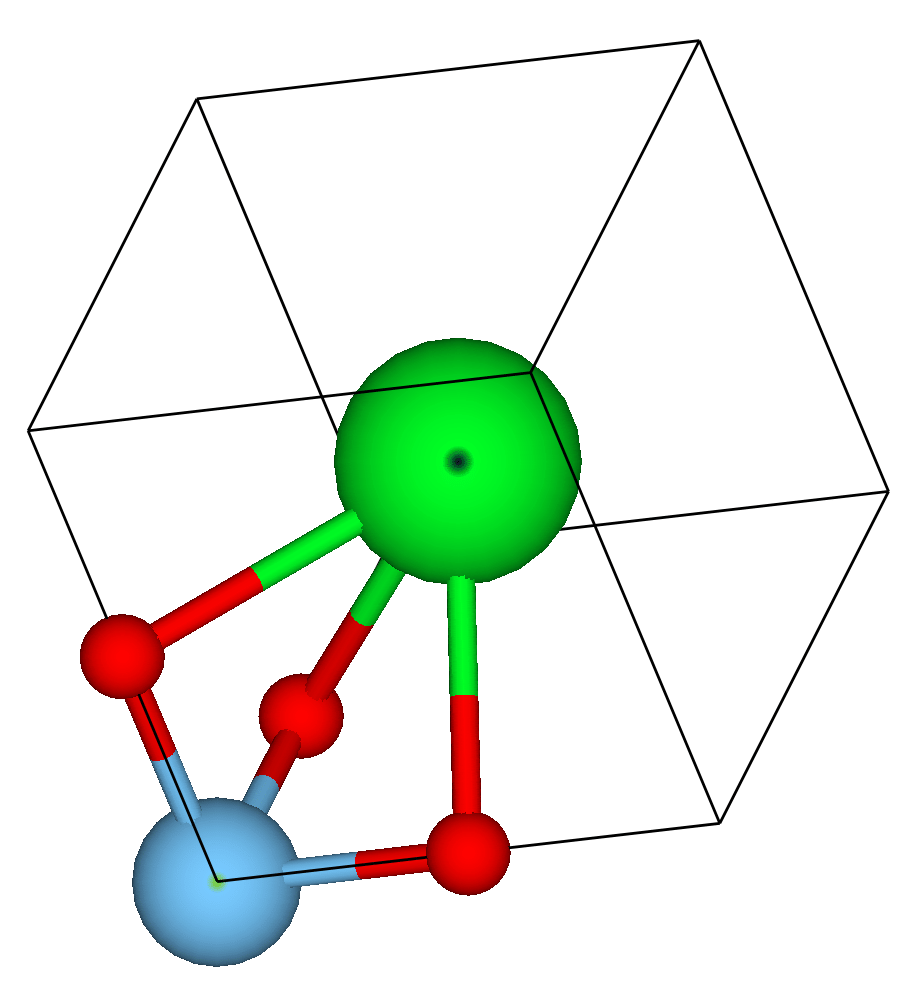}
        \caption{Unit cell}
        \label{fig:sub1}
    \end{subfigure}\hspace{25mm}
    \begin{subfigure}{.25\textwidth}
        \centering
        \includegraphics[width=1.0\linewidth]{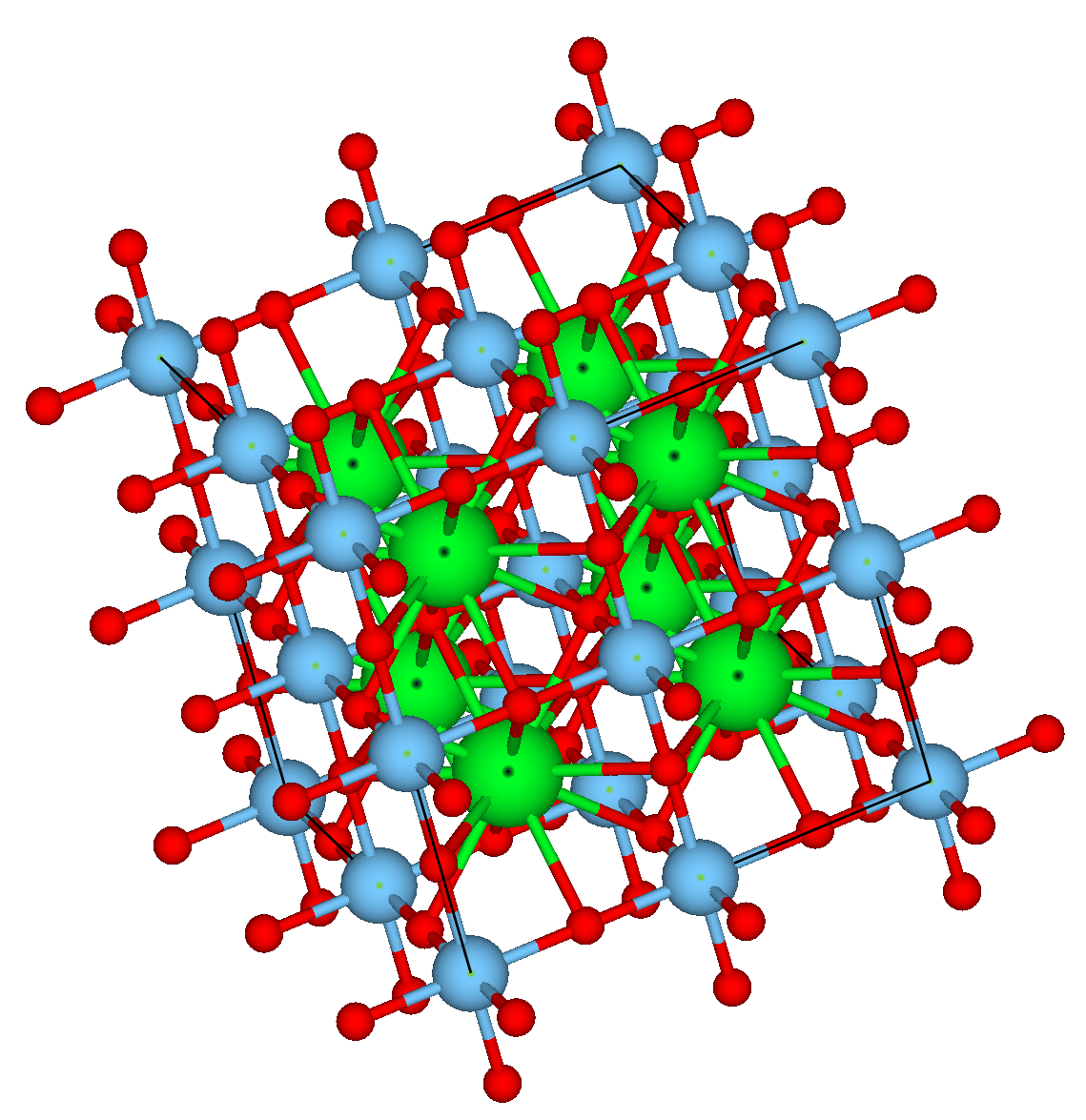}
        \caption{Supercell}
        \label{fig:sub2}
    \end{subfigure}
    \caption{Most stable configuration of \srtio}
    \label{fig:srtio3_2}
\end{figure}

\paragraph{\bf Composition.}
A composition is the chemical formula that describes the ratio of ions that belong 
to the unit cell. The chemical formula contains {\em anions}, negatively charged
ions, and {\em cations}, positively charged ions.
%We will use $C$ to denote a chemistry and $c_i \in \Z^*$ to denote the charge 
%of atom $i \in C$.
The chemical formula is a way of presenting information about the {\em chemical
proportions} of ions that constitute a particular chemical compound, and it 
does not provide any information about the exact number of atoms in the unit cell. 
More formally, the composition is defined by a set of distinct chemical elements 
$\{ e_1, e_2, \dots, e_m\}$, their multiplicity $n_i$, and a non-zero integer charge 
$q_i$ for each element $i$. 
The number $m$ denotes the total number of distinct chemical elements, and 
$n_i/\sum_{j=1}^{m}{n_j}$ is the proportion of the atoms of type $e_i$ in the 
unit cell.
%In addition, each chemical element $e_i$ is associated with a non-zero 
%{\em charge} $q_i \in \zcal^*$.% and a {\em radius} $r_i$.
It is required that the sum of the charges adds up to zero, i.e. , 
$\sum_{i=1}^{m}{q_in_i} = 0$, so that the unit cell is charge neutral.
For example, the composition for Strontium Titanate, \srtio, denotes that the 
following  hold. For every ion of strontium (Sr) in the unit cell, there exists 
one ion of titanium (Ti) and three ions of oxygen (O). Furthermore, the charge 
of  every ion of Sr is $+2$, of Ti is $+4$, and of O is $-2$. Hence, when the ratios 
of the ions are according to the composition, the charge of the unit cell is zero.

Another parameter for every atom is the {\em atomic radius}. This usually 
corresponds to the distance from the center of the nucleus to the boundary of 
the surrounding shells of electrons. Since the boundary is not a well-defined 
physical entity, there are various non-equivalent definitions of atomic radius. 
In crystal structures though, the {\em ionic radius} is used and usually is 
treated as a hard sphere. Thus, we will use $\rho_i$ to denote the ionic radius 
of the element $e_i$.

%Note, both the charge of an ion and its corresponding ionic radius can be found
%in 

\paragraph{\bf Unit cell parameters.}
Unit cell parameters provide a formal description of the parallelepiped that represents 
the unit cell. These include the lengths $y_1, y_2, y_3$ of the
parallelepiped in every dimension and the angles $\theta_{12}$, $\theta_{13}$, 
and $\theta_{23}$ between the corresponding facets. 
For brevity, we denote $y = (y_1, y_2, y_3)$ and  $\theta = (\theta_{12}, 
\theta_{13}, \theta_{23})$, and we use $(y,\theta)$ to denote the unit cell parameters.

\paragraph{\bf Arrangement.}
An arrangement describes the position of each atom of the composition in 
the unit cell. The position of ion $i$ is specified by a point 
$x_i = (x_{i1}, x_{i2}, x_{i3})$ in the parallelepiped defined by the 
unit cell parameters; fractional coordinates $x_i$ denote the location of 
the nucleus of the ion $i$ in the unit cell. A unit cell parameters-arrangement combination 
$(y, \theta, x)$ in a unit cell with $n$ ions is a point in the 
$3n+6$-dimensional space.
%, where $x_{i1} \in [0,y_1]$, $x_{i2} \in [0,y_2]$
%and $x_{i3} \in [0,y_3]$. Formally speaking, $x_i$ own might define a point 
%{\em outside} the parallelepiped for specific . For notation simplicity though, we assume that linear 
%transformation defined by the angles $\theta_{xy}$, $\theta_{xz}$, and 
%$\theta_{yz}$ is applied to the point $f_i$.
For any two points $x_i$ and $x_j$ we will use $\dist(x_i,x_j)$ to denote 
their Euclidean distance.  
%When it is clear from the context, for notation simplicity, we will use 
%$r_{i,j}$ instead.

As we have already said, a unit cell parameters-arrangement configuration $(y,\theta,x)$
defines the period of an infinite structure that covers the whole 3d space.
To get some intuition, assume that we have an orthogonal unit cell, i.e., 
all the angles are 90 degrees. Then for every ion with position 
$(x_{i1},x_{i2},x_{i3})$ in the unit cell, there exist ``copies'' of the ion 
in the positions $(k_1 \cdot y_1+x_{i1}, k_2 \cdot y_2+x_{i2}, k_3 \cdot y_3+x_{i3})$ 
for every possible combination of integers $k_1,k_2$, and $k_3$.
A unit cell parameters-arrangement configuration is {\em feasible} if the hard spheres of any two 
ions of the crystal structure do not overlap; formally, it is feasible if for every two
ions $i$ and $j$ it holds that $\dist(x_i,x_j) \geq \rho_i + \rho_j$.

%%%%%%%%%%%%%%%%%%%
\subsection{Energy}
\label{sec:energy}
Any unit cell parameters-arrangement configuration of a composition corresponds to a {\em 
potential energy}. When the number of ions in the unit cell is fixed,  the
set of configurations define the {\em potential energy surface}. 

%There are several different ways of computing the energy of a configuration that differ both in accuracy and computational cost. 
%{\em Density functional theory} and {\em interatomic forcefields} are two prominent examples for computing the energy of a unit cell parameters-arrangement configuration \cite{parr1980,kohn1965,van1990}. Density functional theory provides more accurate results but is computationally expensive, while interatomic forcefields are less accurate but relatively cheap to compute in practice. 
{\em Buckingham-Coulomb} potential is among the most well adopted 
%interatomic forcefields
methods for computing energy \cite{buckingham1938,walker2010fundamentals} and it is
the sum of the Buckingham potential and the Coulomb potential.
%for every pair of atoms of the crystal structure.
%where at least one of the atoms of the pair is in the unit cell.
The Coulomb potential is {\em long-range} and depends only on the charges and the
distance between the ions; for a pair of ions $i$ and $j$, the Coulomb energy 
is defined by 
\begin{align}
\label{eq:coulomb}
CE(i,j) := \frac{q_iq_j}{\dist(x_i,x_j)}.
\end{align}
Note, ions $i$ and $j$ can be in {\em different} unit cells.

The Buckingham potential is {\em short-range} and depends on the species of the 
ions and their distance. More formally, it depends on  
positive composition-dependent constants $A_{e_i,e_j}, B_{e_i,e_j}$, and
$C_{e_i,e_j}$ for every pair of species $e_i$ and $e_j$; here $i$ can be equal 
to $j$~\footnote{The Buckingham constants are composition-depended since they can 
have small discrepancies in different compositions. For example the constants 
$A_{\text{Ti,O}}, B_{\text{Ti,O}}$, and $C_{\text{Ti,O}}$  for \srtio can be different than those for \mgtio. There is a long 
line of research in computational chemistry that tries to learn/estimate the
Buckingham constants for various compositions. In addition, more than one set of
Buckingham constants can be available for a given composition.}. 
So, for the pair of ions $i$, of specie $e_i$, and $j$, of specie $e_j$, 
%that are at distance $r_{ij} \leq r^*$, 
the Buckingham energy is 
\begin{align}
\label{eq:buckingham}
BE(i,j) := A_{e_i,e_j} \cdot \exp(-B_{e_i,e_j} \cdot \dist(x_i,x_j)) - \frac{C_{e_i,e_j}}{\dist(x_i,x_j)^6}.
\end{align}
Again, ions $i$ and $j$ can be in  different unit cells.

Let $S(x_i,\rho)$ denote the sphere with centre $x_i$ and radius $\rho$.
The total energy of a crystal structure whose unit cell is characterized 
by $n$ ions with arrangement $x = (x_1, \ldots, x_n)$ is then defined
\begin{align*}
%\label{eq:energy}
E(y,\theta, x) = \lim_{\rho \to \infty} \sum_{i=1}^n \sum_{j \neq i, j \in S(x_i, \rho)} \left( BE(i,j)+ CE(i,j) \right).
\end{align*}
$E(y,\theta, x)$ {\em conditionally converges} to a certain value \cite{pickard2018} and 
usually numerical approaches are used to compute it. For this reason, and since 
we aim for an oblivious algorithm, we view the computation of the energy of a
structure as a {\em black box}. More specifically, we assume that we have an 
oracle that given any structure $(y,\theta,x)$, it returns its corresponding 
energy. %We use $\en(y,\theta,x)$ to denote the call to this oracle. 

\begin{question}
\label{que:energy}
Given a composition and Buckingham parameters for it, find a simple, purely
combinatorial way that approximates the energy for every crystal structure.
\end{question}

\begin{question}
Given a composition $\{e_1, e_2, \ldots, e_m\}$ and an oracle that computes the 
energy of every structure for this composition, learn efficiently (with respect to 
the number of oracle calls) the Buckingham parameters $A_{e_i,e_j}, B_{e_i,e_j}$, 
and $C_{e_i,e_j}$ for every $i,j \in [m]$.
\end{question}

\paragraph{\bf Relaxation.} The {\em relaxation} of a crystal structure 
$(y,\theta,x)$ computes a stationary point on the potential energy surface 
by applying gradient descent starting from $(y,\theta,x)$. The relaxation of 
a structure can change {\em both} the arrangement $x$ of the ions in the unit 
cell {\em and} the unit cell parameters $(y, \theta)$ of the unit cell.
% In practice, the software package GULP \argy{give citation} \cite{gale1997gulp,gale2003gulp} is used.
 % as a black box that takes as input a structure and 
We follow a similar approach as we did with the energy and we assume that 
there is an oracle that given a crystal structure $(y,\theta,x)$ it returns 
the relaxed structure. %We denote this as $\rel(y,\theta,x)$.

\begin{question}
Find an alternative, quicker, way to compute an approximate local minimum when:
%under the following constraints:
\begin{itemize}
\item[a)] the unit cell parameters $(y,\theta)$ of the unit cell are fixed;
\item[b)] the arrangement $x$ of the ions is fixed;
\item[c)] both unit cell parameters and arrangement are free.
\end{itemize}
\end{question}

%\begin{question}
%
%\end{question}

\subsection{Crystal Structure Prediction problems}
In crystal structure prediction problems the general goal is to minimize the
energy in the unit cell. There are two kinds of problems we are concerned. 
The first cares only about the value of the energy and the second one cares
for the arrangement and the unit cell parameters that achieve the minimum energy.
From the computational chemistry point of view, both questions are interesting
in their own right. The existence of a unit cell parameters-arrangement that achieves 
lower-than-currently-known energy usually suffices for constructing a new material.
On the other hand, identifying the arrangement and the unit cell parameters of a crystal
structure that achieves the lowest possible energy can help physical scientists to predict
the properties of the material.

\begin{tcolorbox}[title = {\minenergy}]
{\bf Input:} A composition with its corresponding Buckingham constants, a positive 
integer $n$, and a rational $\hat{E}$.\\
{\bf Question:} Is there a crystal structure $(y,\theta,x)$ for the composition with $n$ ions that is neutrally charged and achieves Buckingham-Coulomb 
energy $E(y,\theta,x) < \hat{E}$?
\end{tcolorbox}

\begin{tcolorbox}[title = {\minstructure}]
{\bf Input:} A composition with its corresponding Buckingham constants, a positive 
integer $n$.\\
{\bf Task:} Find a crystal structure $(y, \theta, x)$ for 
the composition with $n$ ions that is neutrally charged and the Buckingham-Coulomb 
energy $E(y, \theta, x)$ is minimized.
\end{tcolorbox}

The second class of problems, the ones that ultimately computational chemists would
like to solve, take as input only the composition and the goal is to construct a unit 
cell, with any number of atoms, such that the {\em average energy per ion} is 
minimized. 

\begin{tcolorbox}[title = {\avgenergy}]
{\bf Input:} A composition with its corresponding Buckingham constants and a 
rational $\hat{E}$.\\
{\bf Question:} Is there a crystal structure for the composition
that is neutrally charged and $\frac{E(y, \theta, x)}{n} < \hat{E}$?
\end{tcolorbox}

\begin{tcolorbox}[title = {\avgstructure}]
{\bf Input:} A composition with its corresponding Buckingham constants.\\
{\bf Task:} Find a crystal structure for the composition that is
neutrally charged and the average Buckingham-Coulomb energy per ion in the unit
cell, $\frac{E(y, \theta, x)}{n}$, is minimized.
\end{tcolorbox}

Although the problems are considered to be intractable~\cite{2019review}, 
only recently the first {\em correct} NP-hardness result was proven for a variant of CSP~\cite{ADGP20}.
However, for the problems presented, there are no correct NP-hardness results in the literature.

\begin{question}
Provide provable lower bounds and upper bounds for the four problems defined above.
\end{question}

\begin{question}
\label{eq:heur}
Construct a heuristic algorithm that works well in practice.
\end{question}

\section{Local Search} 
\label{sec:local_search}

Local search algorithms start from a feasible solution and iteratively obtain better 
solutions. The key concept for the success of such algorithms, is given a feasible
solution, to be able to {\em efficiently find} an improved one. 
%Gradient descent is among the most common local search algorithm. 
Put formally, a local search algorithm is defined by a {\em neighbourhood function}
$N$ and a {\em local rule} $r$. In every iteration, the algorithm does the following.
\begin{itemize}
\item  Has the current best solution $x$.
\item  Computes the neighbourhood $N(x)$.
\item If there is an improved solution $x'$ in $N(x)$, then it updates $x$ according
to the rule $r$, i.e. $x' = r(N(x))$; else it terminates and outputs $x$.
\end{itemize}

The neighborhood $N(x)$ of a solution $x$ consists of all feasible solutions that are 
``close'' in some sense to $x$. The size of the neighborhood can be constant or a 
function of the input. In principle, the larger the size of the neighborhood, the better 
the quality of the locally optimal solutions. However, the downside of choosing large 
neighborhoods is that, in general, it makes each iteration computationally more expensive.
Running time and quality of solutions are competing considerations, and 
the trade off between them can be determined through experimentation. 

We study the following {\em combinatorial} neighborhoods for  Crystal Structure  
Prediction. All of them keep the unit cell parameters fixed and change only the arrangement
$x$ of the $n$ ions. Thus, for notation brevity, we  define the neighbourhoods only 
with respect to the arrangement $x$. 
%In addition, all the neighborhoods contain {\em only} the feasible arrangements.
\begin{enumerate}
    \item {\bf $k$-ion swap.} This neighborhood consists of all feasible arrangements that 
    are produced by swapping the locations of $k$ ions. The size of this neighbourhood is $O(n^k k!)$. 
    \item {\bf $k$-swap.} This neighborhood is parameterized by a discretization step
    $\delta$. Using $\delta$ we discretize  the unit cell and then we perform
    swaps of $k$ ions with the content of every point of the discretization. So, an ion can swap
    positions with another ion, or simply move to another vacant position. Again, we take into
    account only the feasible arrangements of the ions. The size of this neighbourhood is 
    $O(n^kk!/\delta^{3k} )$.
    \item {\bf Axes.} This neighborhood has a parameter $\delta$ and computes the following for 
    every ion $i$. Firstly, for every dimension it computes a plane parallel to the  
    corresponding facet of the unit cell and contains the ion $i$. The intersection of any pair 
    of these planes defines an ``axis''. Then, this axis is discretized according to $\delta$.
    The neighborhood locates the ion to every point on the discretization on the three axes and
    we keep only the arrangements that are feasible. The size of this neighbourhood is $O(n/\delta)$.
%    
%    \begin{figure}[h!]
%      \centering
%        \includegraphics[width=0.4\linewidth]{axis.pdf}
%        \caption{Example of axes neighborhood on a $2$-dimensional square grid with one atom.}
%        \label{fig:axis}
%    \end{figure}
%    \item {\bf Pushing.} This is a modification of the Axes neighborhood which modifies
%    \item {\bf Perturbation.} This is a {\em randomised} neighborhood.
\end{enumerate}

In all of our neighborhoods, we are using a {\em greedy} rule to choose $x'$; $x'$ is an 
arrangement that achieves the minimum energy in $N(x)$.

\section{Algorithms}
\label{sec:algorithms}

We propose two algorithms. The first one is a step towards answering Open Question~\ref{que:energy} while the second is a heuristic for \minstructure problem.

For Open Question~\ref{que:energy}, we propose the {\em depth energy computation} for estimating the energy of a structure. Our algorithm has a single parameter, the depth parameter $k$, and works as follows. Given a crystal structure, it creates $k$ layers around the unit cell with copies of the structure. So, for the unit cell parameters $(y,\theta)$ and the arrangement $x$ of $n$ atoms the energy is $E(y,\theta, x) = \sum_{i=1}^n \sum_{j \neq i, j \in D(k)} \left( BE(i,j)+ CE(i,j) \right)$, where $D(k)$ denotes the set of ions in the $k$ layers of unit cells, and $BE(i,j)$ and $CE(i,j)$ are computed as in Equations~\ref{eq:buckingham} and~\ref{eq:coulomb} respectively.

For \minstructure problem, we slightly modify basin hopping.
In a step of basing hopping, a structure is randomly chosen and it is followed by a relaxation. Our
algorithm applies a combinatorial local search using the Axes neighborhood, since this turned out to
be the best among our heuristics, before the relaxation. 
So, we will perform a relaxation, {\em only after} combinatorial local search cannot further improve 
the solution. Our algorithm can be used as a standalone one and it can also be integrated into {\em any other} heuristic algorithm for the Crystal Structure Prediction problem since it is oblivious. 
In addition, it provides a very fast criterion that when it succeeds it guarantees finding a lower 
energy crystal structure.

\section{Experiments}\label{sec:experiments}
In this section we evaluate our algorithms via experimental simulations. We first focus on \srtio which we use as a benchmark. We do this because it is a well studied composition for which the Crystal Structure Prediction problem is solved.
%In addition, the Buckingham constants for this composition are estimated with confidence and thus during we can evaluate depth energy computation.
We have implemented the algorithms in Python 2.7 and we use the Atomic Simulation Environment (\url{https://wiki.fysik.dtu.dk/ase/}) package for setting up, manipulating, running, visualizing and analyzing atomistic simulations.
All experiments were performed on a 4-core Intel i7-4710MQ with 8GB of RAM.
%All experiments were performed on an 8-core Intel Xeon E5-1620 (3.5GHz) with 16GB of RAM.

%\subsection{Energy Computation}

\iffalse
\begin{center}

\begin{tabular}{ l | c c | c c  } 
    & \multicolumn{2}{c|}{15 atoms}    &   \multicolumn{2}{c}{20 atoms}\\ 
    \hline
    k   & \makecell{Time} & \makecell{Energy difference}   & \makecell{Time} & \makecell{Energy difference}\\
    \hline
    $1$ &     $41$    &   $0.0639$    &   $73$    &   $0.0670$    \\
    $2$ &     $115$   &   $0.0226$    &   $174$   &   $0.0238$    \\ 
    $3$ &     $231$   &   $0.0114$    &   $390$   &   $0.0120$    \\
    $4$ &     $437$   &   $0.0068$    &   $778$   &   $0.0072$    \\
    $5$ &     $757$   &   $0.0045$    &   $1380$  &   $0.0047$    \\
    $6$ &     $1218$  &   $0.0032$    &   $2244$  &   $0.0033$    \\
    \hline
    GULP      &     $20$    &   -           &   $22$    &   -           \\
\end{tabular}
\captionof{table}{Comparison between depth approach and GULP for \srtio. Time is in milliseconds (ms) 
and energy in electronvolts (eV). The energy difference shows the average difference in energy between 
the depth approach and GULP. Results averaged over $2000$ random feasible structures.}
\label{table:depth_time_energy}
\end{center}
\fi

\begin{center}

\begin{tabular*} {\textwidth}{@{\extracolsep{\fill}} c | c c c c c c @{}} 
    & \multicolumn{6}{c}{Energy difference} \\ 
    \hline
    k   & $1$ &  $2$ & $3$ & $4$ & $5$ & $6$\\
    \hline
    $15$ atoms &   $0.0639$  & $0.0226$  & $0.0114$ &  $0.0068$ & $0.0045$    & $0.0032$    \\
    $20$ atoms &    $0.0670$  &   $0.0238$  &  $0.0120$ &   $0.0072$ &   $0.0047$ &   $0.0033$  \\
\end{tabular*}
\captionof{table}{Comparison between depth approach and GULP for \srtio. Energy is in electronvolts (eV). The energy difference shows the average difference in energy between the depth approach and the energy calculated by GULP. Results averaged over $2000$ random feasible structures.}
\label{table:depth_time_energy}
\end{center}

We evaluate the depth energy computation in several different dimensions. For all the experiments we performed for energy computation, we fixed the unit cell to be cubic. Firstly, we evaluate how depth energy computation behaves with respect to $k$. We see that the method converges very fast and $k=6$ already achieves accuracy of three decimal points. Then, we compare our depth approach against GULP; see Table~\ref{table:depth_time_energy}.
%We observe that our naive implementation requires significantly more CPU time than GULP, but this was not our goal~\footnote{GULP is highly parallelised using MPI, while our implementation is sequential.}.
Our goal is to provide an intuitively simpler to interpret and work with method for computing the energy. Even though the energy calculated by the depth approach differs from the one calculated by GULP, we observe that the relative energies between two random arrangements remain usually the same even for $k=1$. To be more precise, let $E_1(x)$ denote the energy of a feasible arrangement $x$ when $k=1$ and let $E_G(x)$ denote the energy of this arrangement as it is computed by GULP. Our experiments show that if for two random feasible arrangements $x_1$ and $x_2$ it holds that $E_1(x_1) <  E_1(x_2)$, then $E_G(x_1) <  E_G(x_2)$ for $99.8\%$ of 1000 pairs of arrangements. This percentage reaches $100 \%$ for $k=6$. For the ``special'' arrangement of ions $x^*$ that minimizes the energy computed by GULP, that is $x^* = \argmin E_G(x)$, our experiments show that it is {\em always true} that $E_k(x^*) < E_k(x)$, for every $k=1, \ldots, 6$, where $x$ is a random feasible structure over $10000$ of them. So, this is a good indication that the arrangement that minimizes the energy for $k=1$, also minimizes the energy overall. We view this as a striking result; it significantly simplifies the problem thus new, analytical, methods can be derived for the problem.

\iffalse

\begin{center}
\begin{tabular}{ |c||c|c| } 
    \hline
    \multicolumn{3}{|c|}{Ordering of energy between pairs of random structures} \\
    \hline
    & 15 atoms    &   20 atoms\\ 
    \hline
    & \makecell{Success rate}   &   \makecell{Success rate}\\ 
    \hline
    Depth $1$ &     $98.9 \%$   &   $99.3 \%$    \\
    Depth $2$ &     $99.8 \%$   &   $99.8 \%$    \\ 
    Depth $3$ &     $99.8 \%$   &   $99.8 \%$    \\
    Depth $4$ &     $99.8 \%$   &   $99.8 \%$    \\
    Depth $5$ &     $99.9 \%$   &   $99.8 \%$    \\
    Depth $6$ &     $100 \%$    &   $99.9 \%$    \\
    \hline
\end{tabular}
\captionof{table}{\srtio. Results averaged over $1000$ pairs of random structures.}
\label{table:ordering}
\end{center}

\begin{center}
\begin{tabular}{ |c||c| } 
    \hline
    \multicolumn{2}{|c|}{Ordering of energy between random structures and global optimum} \\
    \hline
    & \makecell{Success rate} \\ 
    \hline
    Depth $1$ &     $99.99 \%$    \\
    Depth $2$ &     $100.0 \%$    \\ 
    Depth $3$ &     $100.0 \%$    \\
    \hline
\end{tabular}
\captionof{table}{\srtio ($5$ atoms). Results averaged over $10000$ random arrangements.}
\label{table:ordering_2}
\end{center}

\fi

%\subsection{Comparison of the neighborhoods}
The next set of experiments compares the three neighborhoods described in 
Section~\ref{sec:local_search} for \srtio\footnote{The values of the Buckingham parameters can be found in the appendix}. We compare them in several different dimensions: the average 
CPU time they need in order to find a local optimum with respect to their combinatorial 
neighborhood and the average drop in energy until they reach such a local optimum 
(Tables~\ref{table:neighbourhoods_comparison} and \ref{table:neighbourhoods_comparison_ytio}); the average CPU time the relaxation needs 
starting from such local minimum and the average drop in energy from relaxation 
(Tables~\ref{table:neighbourhoods_comparison_gulp} and \ref{table:neighbourhoods_comparison_gulp_ytio}). In addition, for the case of \srtio, we compare how often we can find 
the optimal arrangement from a single structure. 
We observe that the Axes neighborhood has the best tradeoff between energy drop and CPU time.
The $2$-ion-swap neighbourhood outperforms the other two in terms of running time, however it seems to decrease the probability of finding the best arrangement when performing a relaxation on the resulting structures. This renders the use of $2$-ion-swap neighbourhood inappropriate.
Axes neighborhood is significantly faster and performs smoother in terms of running time than the $2$-swap neighbourhood. However the latter one performs better with respect to the energy drop, which is expected since axes is a subset of the 2-swap neighbourhood. In addition, the relaxation from the local minimum found by $2$-swap significantly improves the probability of finding the best arrangement with only one relaxation.
%
%The above experiments highlight the fact that the bigger the size of the neighbourhood, the better the probability of finding the best arrangement.
%Considering the trade-off between efficiency and running time, we have chosen the Axes neighbourhood in our next set of experiments. 
%
We should highlight that there exist structures where the relaxation cannot improve 
their energy, but the neighborhoods do; hence using Axes neighborhood we can escape from some local minima of the continuous space.

\begin{center}
\begin{tabular*}{\textwidth}{@{\extracolsep{\fill}}  l c c c c  @{}} 
    Neighbourhood   &   Running time    &   Time stdev  &   Energy drop &   Energy drop stdev   \\ 
    \hline
    Axes        &   $5.36$  &   $1.54$  &   $13.46$ &   $10.60$ \\ 
    $2$-ion swap&   $0.96$  &   $0.33$  &   $7.75$  &   $8.45$ \\
    $2$-swap    &   $34.66$ &   $14.06$  &   $16.21$ &   $10.94$ \\
\end{tabular*}
\captionof{table}{Comparison of local neighbourhoods for reaching a combinatorial minimum for \srtio with 15 atoms per unit cell and $\delta = 1\angstrom$ ($375$ grid points). 
Time is in seconds and energy in electronvolts (eV). Results averaged over $1000$ arrangements.}
\label{table:neighbourhoods_comparison}
\end{center}

\begin{center}
\begin{tabular*}{\textwidth}{@{\extracolsep{\fill}}  l c c c c c @{}} 
    Neighbourhood  & Running time & \makecell{Time \\ stdev} & Energy drop & \makecell{Energy drop \\ stdev} & \makecell{Global \\ minimum}\\ 
    \hline
    Random structures-GULP &    $8.80$ &    $6.35$ &    $18.60$ &   $10.26$  &   $6.6\%$ \\
    Axes-GULP        &   $7.92$  &   $6.16$  &   $5.53$ &   $2.28$ &   $10.0\%$ \\ 
    $2$-ion swap-GULP  &   $8.82$  &   $6.25$  &   $11.09$  &   $5.64$    & $4.7\%$ \\
    $2$-swap-GULP    &   $5.14$ &   $5.08$  &   $2.79$ &   $1.05$  &   $14.8\%$\\
\end{tabular*}
\captionof{table}{Evaluation of relaxation procedure
after using a combinatorial neighborhood for \srtio with 15 atoms per unit cell. 
Time is in seconds and energy in electronvolts (eV). Results averaged over $1000$ arragnements.}
\label{table:neighbourhoods_comparison_gulp}
\end{center}

\begin{center}
\begin{tabular*}{\textwidth}{@{\extracolsep{\fill}}  l c c c c  @{}} 
    Neighbourhood   &   Running time    &   Time stdev  &   Energy drop &   Energy drop stdev   \\ 
    \hline
    Axes        &   $7.56$  &   $2.08$  &   $8.23$  &   $3.71$\\ 
    $2$-ion swap&   $0.88$  &   $0.43$  &   $1.16$  &   $1.80$ \\
    $2$-swap    &   $27.93$ &   $9.98$  &   $10.26$ &   $4.05$\\
\end{tabular*}
\captionof{table}{Comparison of local neighbourhoods for reaching a combinatorial minimum for \ytio and $\delta = 1\angstrom$ ($343$ grid points). 
Time is in seconds and energy in electronvolts (eV). Results averaged over $1000$ arrangements.}
\label{table:neighbourhoods_comparison_ytio}
\end{center}

\begin{center}
\begin{tabular*}{\textwidth}{@{\extracolsep{\fill}}  l c c c c @{}} 
    Neighbourhood  & Running time & \makecell{Time \\ stdev} & Energy drop & \makecell{Energy drop \\ stdev} \\ 
    \hline
    Random structures-GULP  & $2.81$ & $1.45$ & $12.84$ & $4.88$ \\
    Axes-GULP               & $2.30$ & $1.20$ & $5.09$  & $1.81$ \\ 
    $2$-ion swap-GULP       & $2.47$ & $2.25$ & $12.29$ & $4.38$ \\
    $2$-swap-GULP           & $1.99$ & $1.09$ & $3.11$  & $0.93$ \\
\end{tabular*}
\captionof{table}{Evaluation of relaxation procedure
after using a combinatorial neighborhood for \ytio. 
Time is in seconds and energy in electronvolts (eV). Results averaged over $1000$ arragnements.}
\label{table:neighbourhoods_comparison_gulp_ytio}
\end{center}

%\subsection{Comparison with basin hopping}

Next, we compare our algorithm for \minstructure against basin hopping where the next structure to relax is chosen at random. Based on the results of our previous experiments, we have chosen the Axes neighbourhood as an intermediate step before the relaxation. We have run these algorithms $200$ times for \srtio with $15$ atoms per unit cell, and $25$ times for \srtio with $20$ atoms per unit cell.
We report how the energy varies with respect to time until the best arrangement is found (Fig. \ref{fig:srtio3}) and we report other statistics that further validate our approach (Table \ref{table:statistics}).
%This time we leave the unit cell parameters free as well.
As we can see, it is relatively easy to reach low levels of energy and the majority of time is needed to find the absolute minimum.
In addition, the overhead posed by the use of the neighbourhood search divided by the time needed by the basin hopping to find the global minimum decreases as the number of the atoms in the unit cell increases.

\begin{center}
\begin{small}
\begin{tabular*}{\textwidth}{@{\extracolsep{\fill}} l c c c c c c@{}} 
    Algorithm  & \makecell{Number of \\ atoms}  & \makecell{Total time \\ mean} & \makecell{Total time \\ stdev} & Relaxations & \makecell{Time for \\ relaxations} & \makecell{Time for \\ local search} \\ 
    \hline
    Axes-GULP   & $15$ &  $227.89$  & $287.21$ &  $13.24$    & $126.26$ &   $101.63$ \\
                & $20$ &  $2280.57$  & $781.66$ &  $104.33$    & $1016.72$ &   $1049.13$ \\
    Basin hopping &  $15$ &  $167.89$  & $114.89$ &  $18.14$ &   $160.79$ &   $-$ \\
                  &  $20$ &  $5766.20$  & $4748.33$ &  $450.66$ &   $4895.60$ &   $-$ \\
\end{tabular*}
\captionof{table}{Statistics from the experiments depicted in Figures \ref{fig:srtio3} and \ref{fig:srtio3_20} (\srtio with $15$ and $20$ atoms per unit cell).}
\label{table:statistics}
\end{small}
\end{center}

\begin{figure}[h]
    \centering
    \begin{subfigure}{.48\textwidth}
        \centering
        \includegraphics[width=1.0\linewidth]{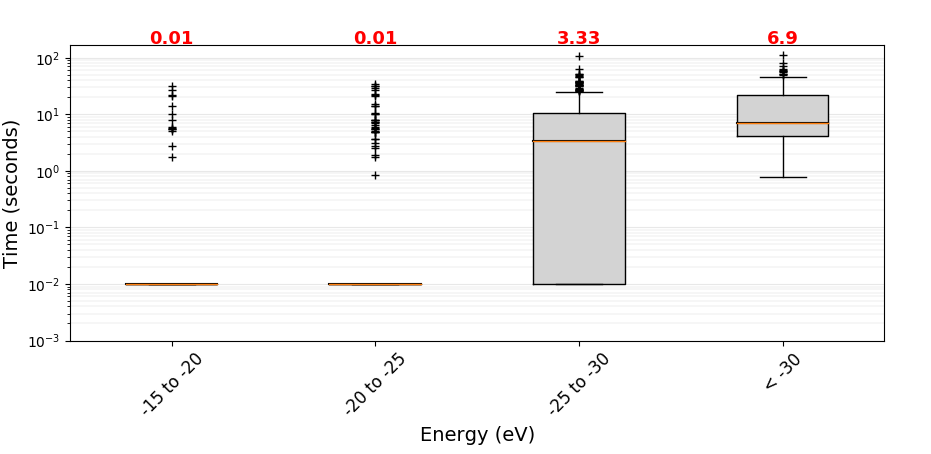}
        \caption{Axes - GULP coarse}
        \label{fig:algorithm_axes_gulp_high_energy}
    \end{subfigure}\hspace{0mm}
    \begin{subfigure}{.48\textwidth}
        \centering
        \includegraphics[width=1.0\linewidth]{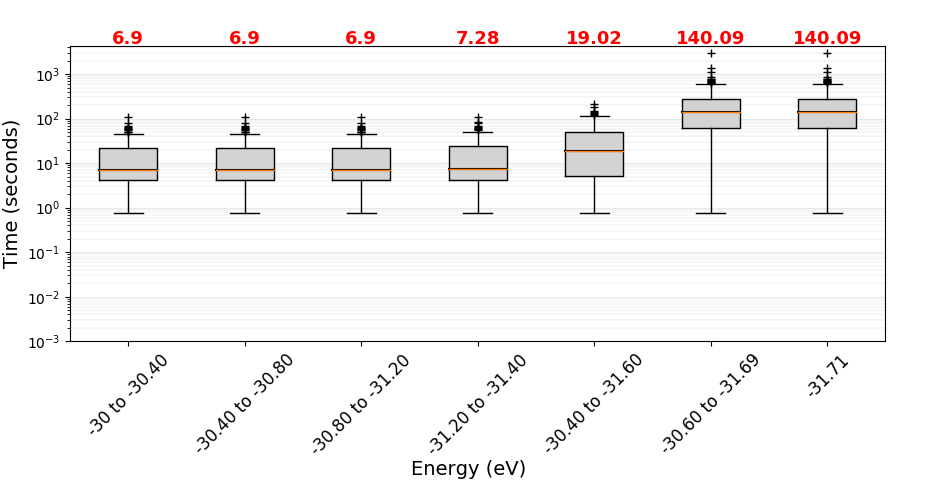}
        \caption{Axes - GULP fine}
        \label{fig:algorithm_axes_gulp_low_energy}
    \end{subfigure}
    
    \begin{subfigure}{.48\textwidth}
        \centering
        \includegraphics[width=1.0\linewidth]{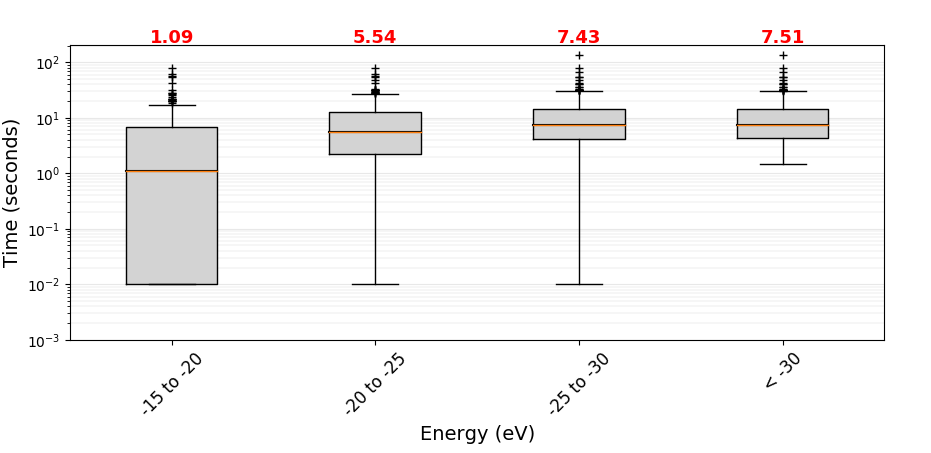}
        \caption{GULP coarse}
        \label{fig:algorithm_gulp_high_energy}
    \end{subfigure}
    \begin{subfigure}{.48\textwidth}
        \centering
        \includegraphics[width=1.0\linewidth]{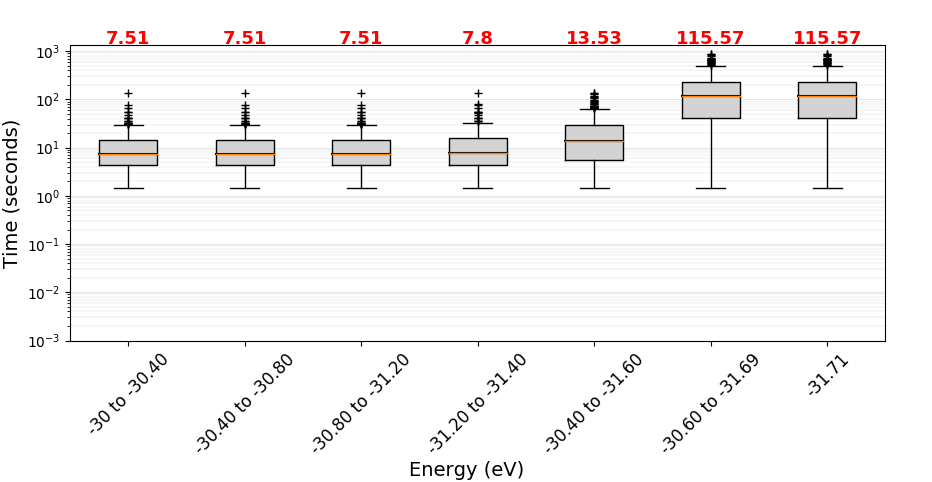}
        \caption{GULP fine}
        \label{fig:algorithm_gulp_low_energy}
    \end{subfigure}
    \caption{Time to reach specific energy levels for \srtio (15 atoms). Figures (a) and (b) correspond to the algorithm of Section \ref{sec:algorithms}. Figures (c) and (d) correspond to basin hopping. The median times needed to reach every energy level are depicted in red on the top of each plot.}
    \label{fig:srtio3}
\end{figure}

\begin{figure}
  \includegraphics[width=\linewidth]{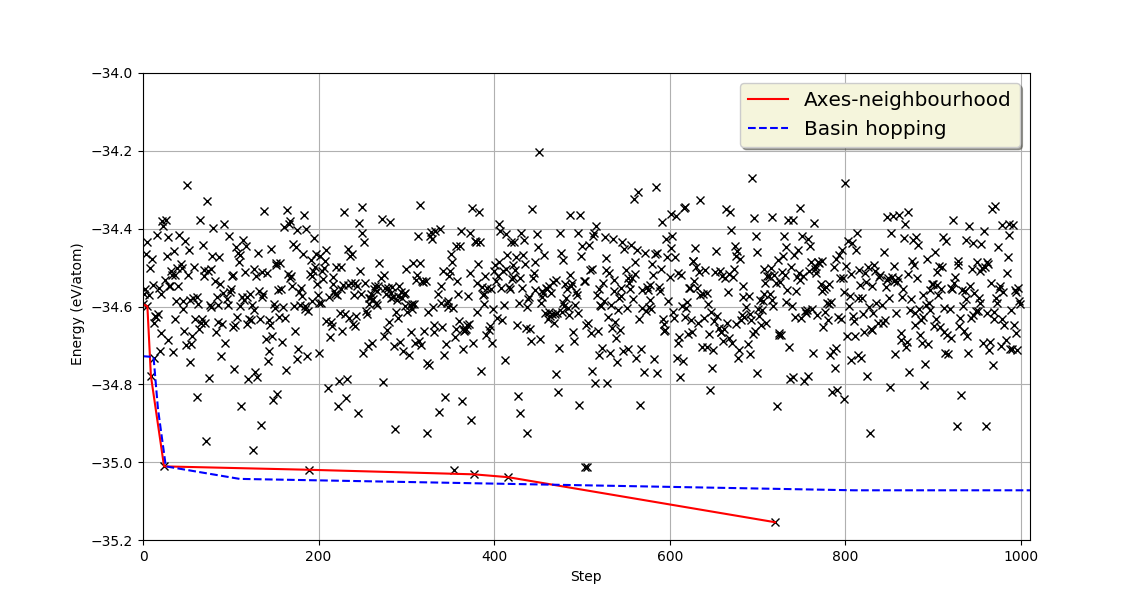}
  \caption{Performance of the Axes algorithm for \ytio. The black points correspond to the energy found after the relaxation of a point computed by the Axes neighborhood at each step. The red line is the lower envelope of the energy found by our algorithm while the blue line corresponds to the lower envelope of basin hopping.}
  \label{fig:energy_plot_axes}
\end{figure}

In our last set of experiments, we compare our algorithm against basin hopping algorithm for \ytio which contains $22$ atoms in its primitive unit cell.
In this set of experiments, we produced $3500$ random structures. We simulated basin hopping by sequentially relaxing the constructed structures. However, none of the relaxations managed to find the optimal configuration.  Our algorithm, using the same order of structures as before, first used the Axes neighborhood as an intermediate step followed by a relaxation; it managed to find the optimal configuration after visiting only 720 structures (Fig. \ref{fig:energy_plot_axes}).
%The execution of these algorithms is depicted in Figure \ref{fig:energy_plot_axes} and \ref{fig:energy_plot_basin}.

\iffalse

\begin{figure}[H]
    \centering
    \begin{subfigure}{.32\textwidth}
        \centering
        \includegraphics[width=1.0\linewidth]{images/axes_structure-1.png}
        \caption{Raw structure}
        \label{fig:algorithm_axes_gulp_high_energy}
    \end{subfigure}\hspace{0mm}
    \begin{subfigure}{.32\textwidth}
        \centering
        \includegraphics[width=1.0\linewidth]{images/FINDSYM_structure-1.png}
        \caption{FINDSYM structure}
        \label{fig:algorithm_axes_gulp_low_energy}
    \end{subfigure}
    \begin{subfigure}{.32\textwidth}
        \centering
        \includegraphics[width=1.0\linewidth]{images/ICSD_structure-1.png}
        \caption{Reported structure}
        \label{fig:algorithm_axes_gulp_low_energy}
    \end{subfigure}
    \caption{(a) The raw structure obtained in step $720$. (b) The structure obtained after passing the structure in (a) through FINDSYM. (c) The experimental crystal structure for \ytio obtained from ICSD. Atoms are coloured as follows: Y (green), Ti (blue) and O (red)}
    \label{fig:srtio3}
\end{figure}

\fi

%The Buckingham potential parameters that were used for both \srtio and \ytio are presented in Table \ref{table:buckingham}.

\section{Conclusions}
In this paper we have introduced and studied the Crystal Structure Prediction problem through the lens
of computer science. This is an important and very exciting problem in computational chemistry, 
which computer scientists are not actively studying  yet. We have identified several open questions 
whose solution  would have significant impact to the discovery of new materials. These problems are 
challenging and several different techniques and machineries from computer science could be applied 
for solving them. Our simple-to-understand algorithms are a first step towards their solution. We hope 
that our algorithms will be used as benchmarks in the future, since more sophisticated techniques for 
basin hopping can be invented. For the energy computation via the depth approach, we conjecture that 
the arrangement that minimizes the energy for $k=1$ or $k=2$, matches the arrangement that minimizes the 
energy when it is computed via GULP. Our numerical simulations provide significant evidence towards this. 
A formal result of this would greatly simplify the objective function of the optimization problem and it 
would give more hope to faster methods for relaxation. In addition, it could provide the foundations for
new techniques for crystal structure prediction. Our algorithm that utilizes the Axes neighborhood as an 
intermediate step before relaxation, seems to speed up the time the standard basin hopping needs to find 
the global minimum. Are there any other neighborhoods that outperform the Axes one? Can local search, or
Axes neighborhood in particular, improve existing methods for crystal structure prediction by a simple
integration as an intermediate step? We believe that this is indeed the case.

\newpage

\bibliographystyle{abbrv}
{\normalsize \bibliography{bibliography}}

\begin{thebibliography}{10}

\bibitem{ADGP20}
D.~Adamson, A.~Deligkas, V.~V. Gusev, and I.~Potapov.
\newblock On the hardness of energy minimisation for crystal structure
  prediction.
\newblock In {\em {SOFSEM} 2020: Theory and Practice of Computer Science - 46th
  International Conference on Current Trends in Theory and Practice of
  Informatics, {SOFSEM} 2020, Limassol, Cyprus, January 20-24, 2020,
  Proceedings}, pages 587--596, 2020.

\bibitem{buckingham1938}
R.~A. Buckingham.
\newblock The classical equation of state of gaseous helium, neon and argon.
\newblock {\em Proceedings of the Royal Society of London. Series A.
  Mathematical and Physical Sciences}, 168(933):264--283, 1938.

\bibitem{call2007}
S.~T. Call, D.~Y. Zubarev, and A.~I. Boldyrev.
\newblock Global minimum structure searches via particle swarm optimization.
\newblock {\em Journal of computational chemistry}, 28(7):1177--1186, 2007.

\bibitem{collins2018}
C.~Collins, G.~Darling, and M.~Rosseinsky.
\newblock The flexible unit structure engine (fuse) for probe structure-based
  composition prediction.
\newblock {\em Faraday discussions}, 211:117--131, 2018.

\bibitem{collins2017}
C.~Collins, M.~Dyer, M.~Pitcher, G.~Whitehead, M.~Zanella, P.~Mandal,
  J.~Claridge, G.~Darling, and M.~Rosseinsky.
\newblock Accelerated discovery of two crystal structure types in a complex
  inorganic phase field.
\newblock {\em Nature}, 546(7657):280, 2017.

\bibitem{curtarolo2013}
S.~Curtarolo, G.~L. Hart, M.~B. Nardelli, N.~Mingo, S.~Sanvito, and O.~Levy.
\newblock The high-throughput highway to computational materials design.
\newblock {\em Nature materials}, 12(3):191, 2013.

\bibitem{deaven1995}
D.~M. Deaven and K.-M. Ho.
\newblock Molecular geometry optimization with a genetic algorithm.
\newblock {\em Physical review letters}, 75(2):288, 1995.

\bibitem{freeman1993}
C.~Freeman, J.~Newsam, S.~Levine, and C.~Catlow.
\newblock Inorganic crystal structure prediction using simplified potentials
  and experimental unit cells: application to the polymorphs of titanium
  dioxide.
\newblock {\em Journal of Materials Chemistry}, 3(5):531--535, 1993.

\bibitem{gale2003gulp}
J.~D. Gale and A.~L. Rohl.
\newblock The general utility lattice program (gulp).
\newblock {\em Molecular Simulation}, 29(5):291--341, 2003.

\bibitem{gautier2015}
R.~Gautier, X.~Zhang, L.~Hu, L.~Yu, Y.~Lin, T.~O. Sunde, D.~Chon, K.~R.
  Poeppelmeier, and A.~Zunger.
\newblock Prediction and accelerated laboratory discovery of previously unknown
  18-electron abx compounds.
\newblock {\em Nature chemistry}, 7(4):308, 2015.

\bibitem{goedecker2004}
S.~Goedecker.
\newblock Minima hopping: An efficient search method for the global minimum of
  the potential energy surface of complex molecular systems.
\newblock {\em The Journal of chemical physics}, 120(21):9911--9917, 2004.

\bibitem{hautier2010}
G.~Hautier, C.~Fischer, V.~Ehrlacher, A.~Jain, and G.~Ceder.
\newblock Data mined ionic substitutions for the discovery of new compounds.
\newblock {\em Inorganic chemistry}, 50(2):656--663, 2010.

\bibitem{lonie2011}
D.~C. Lonie and E.~Zurek.
\newblock Xtalopt: An open-source evolutionary algorithm for crystal structure
  prediction.
\newblock {\em Computer Physics Communications}, 182(2):372--387, 2011.

\bibitem{nosengo2016}
N.~Nosengo.
\newblock Can artificial intelligence create the next wonder material?
\newblock {\em Nature News}, 533(7601):22, 2016.

\bibitem{oganov2006}
A.~R. Oganov and C.~W. Glass.
\newblock Crystal structure prediction using ab initio evolutionary techniques:
  Principles and applications.
\newblock {\em The Journal of chemical physics}, 124(24):244704, 2006.

\bibitem{2019review}
A.~R. Oganov, C.~J. Pickard, Q.~Zhu, and R.~J. Needs.
\newblock Structure prediction drives materials discovery.
\newblock {\em Nature Reviews Materials}, page~1, 2019.

\bibitem{pannetier1990}
J.~Pannetier, J.~Bassas-Alsina, J.~Rodriguez-Carvajal, and V.~Caignaert.
\newblock Prediction of crystal structures from crystal chemistry rules by
  simulated annealing.
\newblock {\em Nature}, 346(6282):343, 1990.

\bibitem{pickard2018}
C.~J. Pickard.
\newblock Real-space pairwise electrostatic summation in a uniform neutralizing
  background.
\newblock {\em Physical Review Materials}, 2(1):013806, 2018.

\bibitem{pickard2006}
C.~J. Pickard and R.~Needs.
\newblock High-pressure phases of silane.
\newblock {\em Physical Review Letters}, 97(4):045504, 2006.

\bibitem{pickard2011}
C.~J. Pickard and R.~Needs.
\newblock Ab initio random structure searching.
\newblock {\em Journal of Physics: Condensed Matter}, 23(5):053201, 2011.

\bibitem{schmidt1996}
M.~U. Schmidt and U.~Englert.
\newblock Prediction of crystal structures.
\newblock {\em Journal of the Chemical Society, Dalton Transactions},
  (10):2077--2082, 1996.

\bibitem{schon2014}
J.~C. Sch{\"o}n.
\newblock How can databases assist with the prediction of chemical compounds?
\newblock {\em Zeitschrift f{\"u}r anorganische und allgemeine Chemie},
  640(14):2717--2726, 2014.

\bibitem{schon1996}
J.~C. Sch{\"o}n and M.~Jansen.
\newblock First step towards planning of syntheses in solid-state chemistry:
  determination of promising structure candidates by global optimization.
\newblock {\em Angewandte Chemie International Edition in English},
  35(12):1286--1304, 1996.

\bibitem{combining_magnets}
T.~Siegrist and T.~A. Vanderah.
\newblock Combining magnets and dielectrics: Crystal chemistry in the bao-
  fe2o3- tio2 system.
\newblock {\em European Journal of Inorganic Chemistry}, 2003(8):1483--1501,
  2003.

\bibitem{magnetic_oxides}
T.~Vanderah, J.~Loezos, and R.~Roth.
\newblock Magnetic dielectric oxides: subsolidus phase relations in the bao:
  Fe2o3: Tio2system.
\newblock {\em Journal of Solid State Chemistry}, 121(1):38--50, 1996.

\bibitem{villars2001}
P.~Villars, K.~Brandenburg, M.~Berndt, S.~LeClair, A.~Jackson, Y.-H. Pao,
  B.~Igelnik, M.~Oxley, B.~Bakshi, P.~Chen, et~al.
\newblock Binary, ternary and quaternary compound former/nonformer prediction
  via mendeleev number.
\newblock {\em Journal of alloys and compounds}, 317:26--38, 2001.

\bibitem{wales1997}
D.~J. Wales and J.~P. Doye.
\newblock Global optimization by basin-hopping and the lowest energy structures
  of lennard-jones clusters containing up to 110 atoms.
\newblock {\em The Journal of Physical Chemistry A}, 101(28):5111--5116, 1997.

\bibitem{walker2010fundamentals}
J.~D. Walker.
\newblock {\em Fundamentals of physics extended}.
\newblock Wiley, 2010.

\bibitem{wang2010}
Y.~Wang, J.~Lv, L.~Zhu, and Y.~Ma.
\newblock Crystal structure prediction via particle-swarm optimization.
\newblock {\em Physical Review B}, 82(9):094116, 2010.

\end{thebibliography}

\newpage
\appendix
\section{Appendix}

\begin{figure}[h]
    \centering
    \begin{subfigure}{.48\textwidth}
        \centering
        \includegraphics[width=1.0\linewidth]{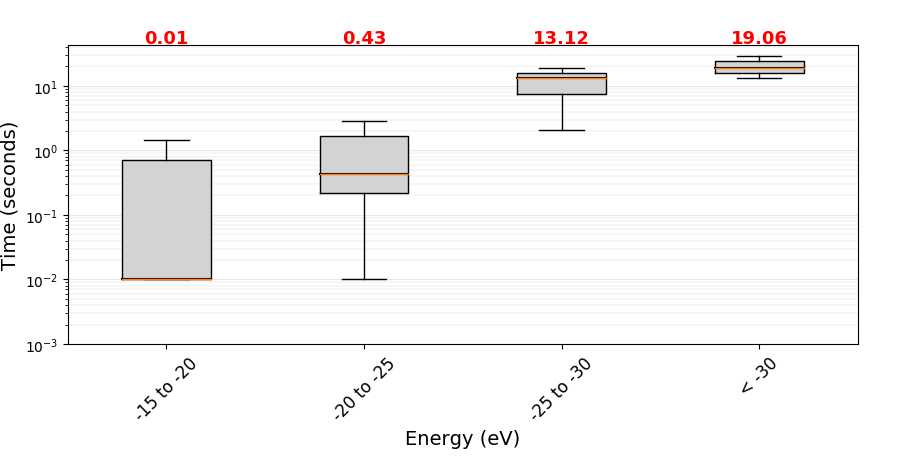}
        \caption{Axes - GULP coarse}
        \label{fig:algorithm_axes_gulp_high_energy_2}
    \end{subfigure}\hspace{0mm}
    \begin{subfigure}{.48\textwidth}
        \centering
        \includegraphics[width=1.0\linewidth]{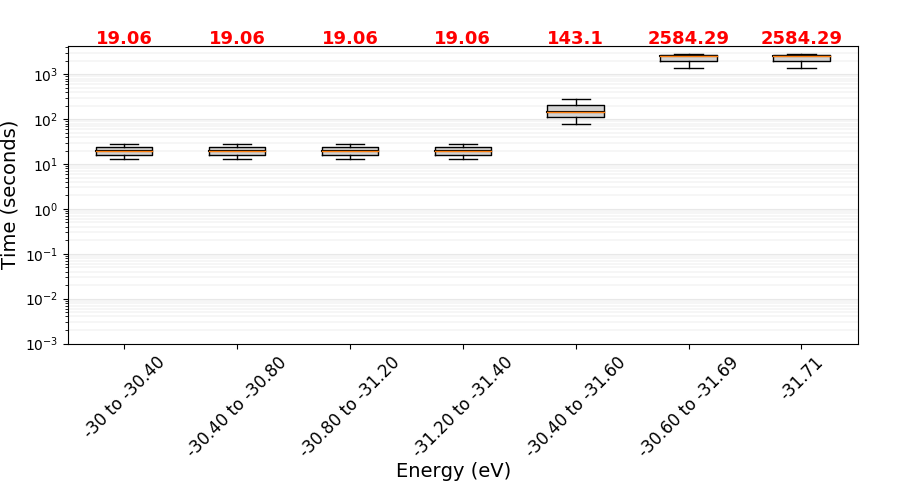}
        \caption{Axes - GULP fine}
        \label{fig:algorithm_axes_gulp_low_energy_2}
    \end{subfigure}
    
    \begin{subfigure}{.48\textwidth}
        \centering
        \includegraphics[width=1.0\linewidth]{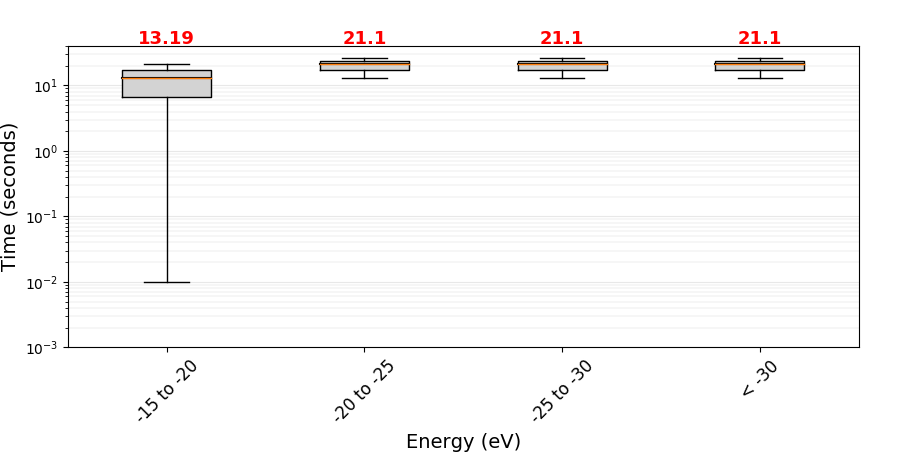}
        \caption{GULP coarse}
        \label{fig:algorithm_gulp_high_energy_2}
    \end{subfigure}
    \begin{subfigure}{.48\textwidth}
        \centering
        \includegraphics[width=1.0\linewidth]{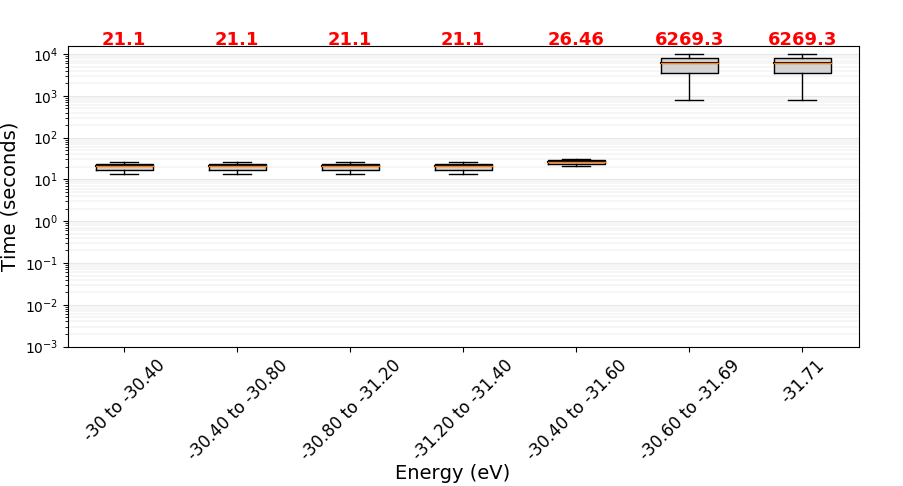}
        \caption{GULP fine}
        \label{fig:algorithm_gulp_low_energy_2}
    \end{subfigure}
    \caption{Time to reach specific energy levels for \srtio (20 atoms). Figures (a) and (b) correspond to the algorithm of Section \ref{sec:algorithms}. Figures (c) and (d) correspond to basin hopping. The median times needed to reach every energy level are depicted in red on the top of each plot.}
    \label{fig:srtio3_20}
\end{figure}

%\subsection{Relaxations for \ytio}

\begin{figure}[h!]
  \includegraphics[width=\linewidth]{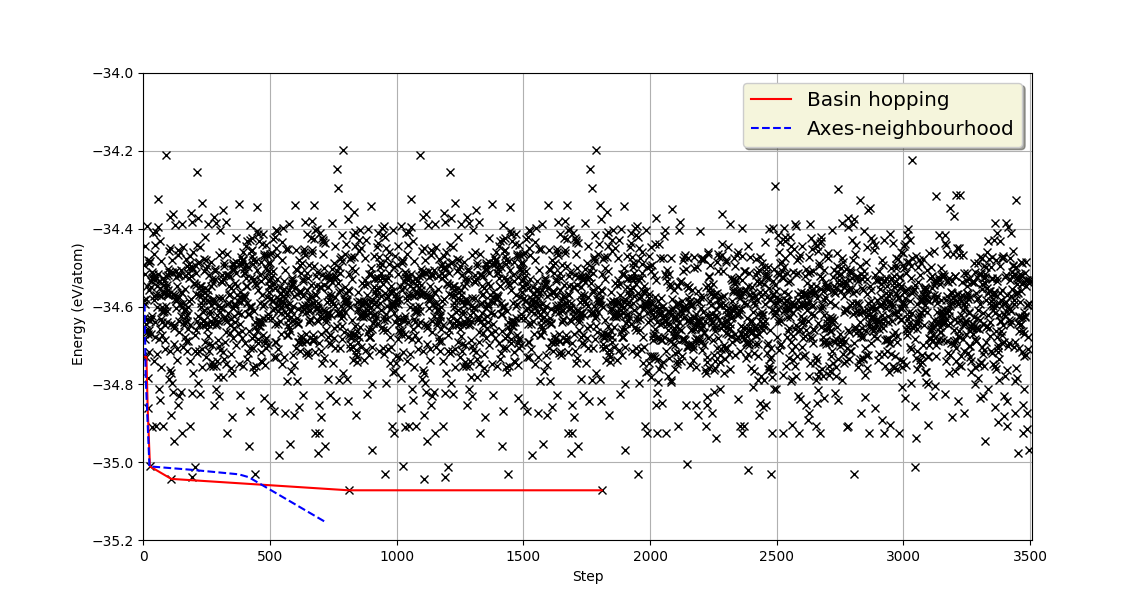}
  \caption{Performance of the basin hopping algorithm for \ytio. The black points correspond to the energy found after the relaxation of a random feasible configuration of atoms at each step. The red line is the lower envelope of the energy found by basin hopping, while the blue line corresponds to the lower envelope of our algorithm.}
  \label{fig:energy_plot_basin}
\end{figure}

%\subsection{Buckingham parameters}
\begin{center}
\begin{small}
\begin{tabular*}{\textwidth}{@{\extracolsep{\fill}} l c c c@{}} 
    Interaction  & $A$ (eV) & $\rho \; (\angstrom)$ & C (eV $\angstrom^{-6}$) \\
    \hline
    $\text{O}^{2-} - \text{O}^{2-}$     &   $1388.77$  & $0.36262$ &  $175$ \\
    $\text{Y}^{3+} - \text{O}^{2-}$ &   $23000$  & $0.24203$ &  $0$\\
    $\text{Sr}^{2+} - \text{O}^{2-}$ &   $1952.39$  & $0.33685$ &  $19.22$\\
    $\text{Ti}^{4+} - \text{O}^{2-}$ &   $4590.7279$  & $0.261$ &  $0$\\
\end{tabular*}
\captionof{table}{The values of Buckingham potential parameters we used in our experiments as they were found in~\cite{collins2018}. All the missing parameters are set to zero.}
\label{table:buckingham}
\end{small}
\end{center}
\end{document}